\documentclass[aps,preprint]{revtex4}

\usepackage{graphicx}

\begin{document}
\title{High photo-excited carrier multiplication by charged InAs dots in AlAs/GaAs/AlAs resonant tunneling diode}

\author{Wangping Wang, Ying Hou, Dayuan Xiong, Ning Li, Wei Lu}
 \email{luwei@mail.sitp.ac.cn}
  \affiliation{National Laboratory for Infrared Physics, Shanghai Institute of Technical Physics, Chinese Academy of Sciences, Shanghai 200083, China}
\author{Wenxing Wang, Hong Chen, Junming Zhou}
 \affiliation{Institute of Physics, Chinese Academy of Sciences, Beijing, China}
\author{E Wu, Heping Zeng}
 \affiliation{State Key Laboratory of Precision Spectroscopy, and Department of Physics, East China Normal University, Shanghai 200062, China}

\begin{abstract}
We present an approach for the highly sensitive photon detection based on the quantum dots (QDs) operating at temperature of 77K. The detection structure is based on an AlAs/GaAs/AlAs double barrier resonant tunneling diode combined with a layer of self-assembled InAs QDs (QD-RTD). A photon rate of 115 photons per second had induced 10nA photocurrent in this structure, corresponding to the photo-excited carrier multiplication factor of $10^{7}$. This high multiplication factor
is achieved by the quantum dot induced memory effect and the resonant tunneling tuning effect of QD-RTD structure.

\end{abstract}

\maketitle

There is currently great interest in exploring highly sensitive photon detection methodologies for potential use in remote sensing, spectroscopy, and even quantum information.\cite{quantuminfo} For the very high sensitivity, the photon detection will be in the photon counting mode, even in single photon counting mode. In order to approach the photon counting mode, very high photo-excited carrier multiplication factor is a basic requirement. Recently it has been demonstrated that structure of resonant tunneling diodes containing a layer of self-assembled quantum dots (QD-RTD) may be used as a very high photo-excited carrier multiplication device under the forward bias\cite{forwardbias} at liquid helium temperature for single photon detection with the effective multiplication factor in the order of $10^{8}$.\cite{PRLphoton} This very high multiplication factor is caused by the storage effect of photo-excited hole in QD near RTD structure. This storage effect can also be achieved by the electrical bias, which presents a pronounced memory effect in current-voltage (\textit{I-V}) characteristics of QD-RTD device.\cite{memory1, memory2} In this letter, we report that the memory phenomenon, induced by the electrical bias, can be used to enhance the photo-excited carrier multiplication factor in QD-RTD structure. The tunneling current passing through the QDs has been used in the multiplication process, so that the operating temperature has been increased from liquid helium (4.2K) in early report\cite{PRLphoton, APLphoton, APLphoton2} to liquid nitrogen (77K) in present work.

\begin{figure} 
  \centering 
  \includegraphics[width=\textwidth]{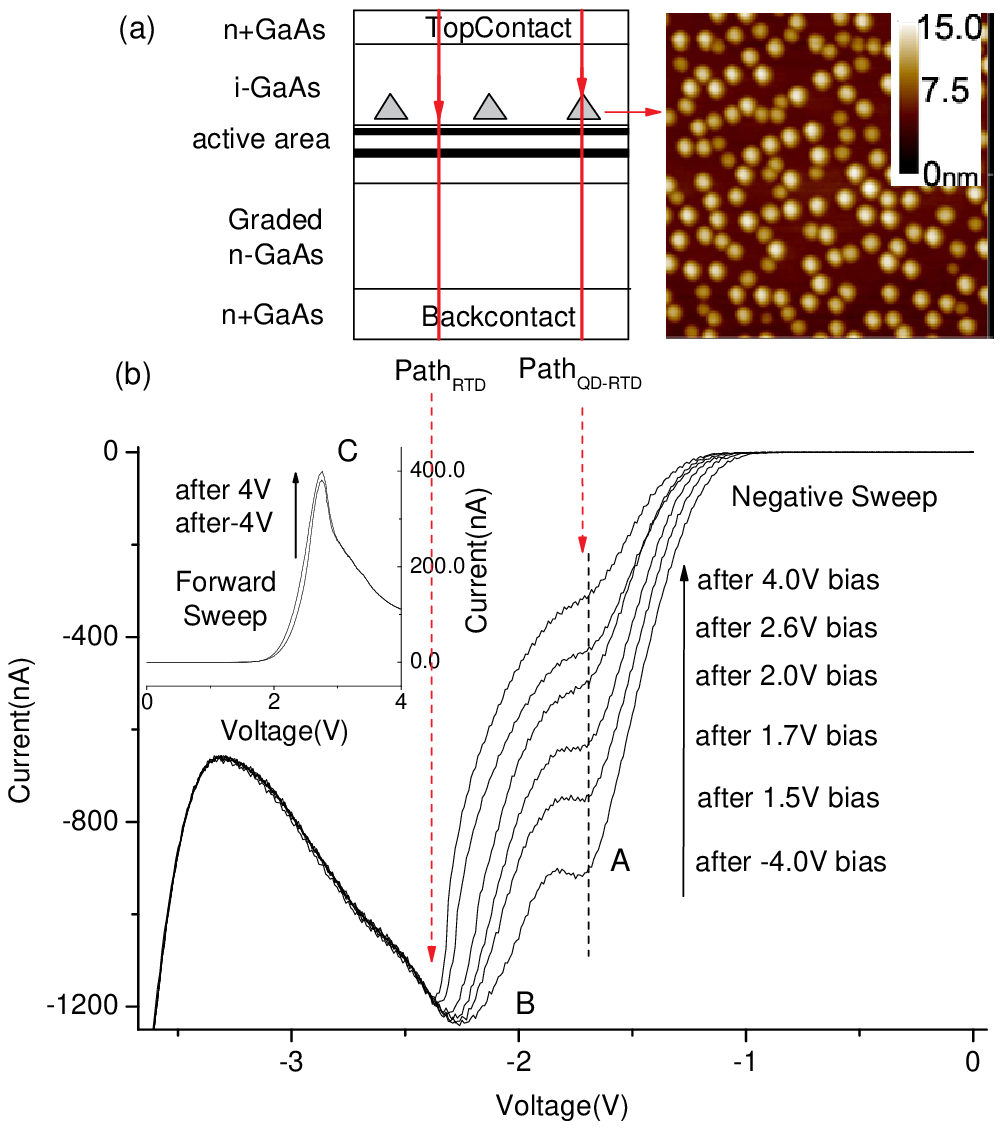}
  \caption{}
  \label{fig:AFMstructureIV} 
\end{figure}

The samples in our experiment were grown by molecular beam epitaxy (MBE) on semi-insulated GaAs (100) substrate. The material layers, from top to bottom, were piled as: a 50nm highly n-doped GaAs ($2\times 10^{18}cm^{-3}$) as top contact layer, a 150nm i-GaAs absorber layer, a layer of self-assembled InAs quantum dots capped by 10nm GaAs, a 2nm GaAs spacer layer, a 11ML AlAs barrier, a 8nm GaAs quantum well, a 11ML AlAs barrier, a 20nm GaAs spacer layer, a 430nm graded n-doped GaAs as bottom contact layer (from $1\times 10^{16}$ to $1\times10^{18}cm^{-3}$), a 15nm AlAs etch-stop layer, a 400nm GaAs buffer layer, and finally a semi-insulating GaAs (100) substrate. In order to investigate the density of QDs, another QD layer of same growing condition was deposited at the surface. The surface morphology was characterized by atomic force microscopy (AFM) as shown in Fig. \ref{fig:AFMstructureIV}(a). The density of QDs is about $5.7\times10^{10}cm^{-2}$, with the QDs' size of 20$\sim$28nm in lateral and 8nm in height.

The resonant tunneling devices with QDs were in a cross-wire geometry fabricated by the standard photolithography and selective etching process.\cite{APLProcess}
 In this geometry, both the collector and the emitter contacts were well separated in different etched wires, while the active area for photo-detection and photo-excited carrier multiplication was formed by the intersection of the two wires. The active area is about $1\mu m\times5\mu m$, with a top contact line (see Fig. \ref{fig:AFMstructureIV}(a)) about 1$\mu m$ wide.

In the measurement, the sample was mounted in a liquid nitrogen Dewar and cooled to 77k. To evaluate the light intensity incident on the QD-RTD device, a continuous wave green laser beam at 532 nm is attenuated continually to tens of photons per second per $\mu m^{2}$ calibrated by a si-avalanche photodiode in single-photon counting regime (PerkinElmer SPCM-AQR). With this laser beam the photon detection sensitivity and the multiplication factor of photo-excited carrier is obtained for our QD-RTD device. The current-voltage (\textit{I-V}) measurement was preformed with a Keithley 236 Programmable Source Measure Unit equipment in the voltage range from -4 to +4V.

According to the early work related with the structure of QD-RTD, the charging condition of QD is the key factor to affect the singularity of electrical and opto-electronic properties.\cite{PRLphoton, APLphoton, APLphoton2} The electron charging effect of QD in our QD-RTD structure is clearly observed in the memory phenomenon as shown in Fig. \ref{fig:AFMstructureIV}(b). There are three peaks (A, B and C) in the \textit{I-V} curves. Peaks B and C are at -2.3V and +2.5V, showing as almost symmetrical counterpart. This is the typical property of resonant tunneling current peak behavior in AlAs/GaAs/AlAs double barrier structure. We attribute the peak B (in Fig. \ref{fig:AFMstructureIV}(b)) and peak C (in the inset of Fig. \ref{fig:AFMstructureIV}(b)) as the normal resonant tunneling peak in double barrier structure of AlAs/GaAs/AlAs with the current through Path$_{\textrm{RTD}}$ shown in Fig. \ref{fig:AFMstructureIV}(a). The unsymmetrical distribution of the QDs beside the AlAs/GaAs/AlAs double barrier structure results in the small difference of the absolute value of voltage for the peak B and C. In contrast with the peak B or C, the peak A clearly observed in Fig. \ref{fig:AFMstructureIV}(b) at -1.7V  has no counterpart peak in forward bias part of \textit{I-V} curve. This peak is weak and at small bias compared with peak B. This peak could be explained by the resonant tunneling current through QD along {Path$_{\textrm{QD-RTD}}$} shown in Fig.  \ref{fig:AFMstructureIV}(a), similarly with the experimental observation in other works.\cite{memory1, memory2} The peak A is gradually smeared out with the ascending of former forward bias which is used to electrically charge the QDs as discussed later.

\begin{figure} 
  \centering 
  \includegraphics[width=\textwidth]{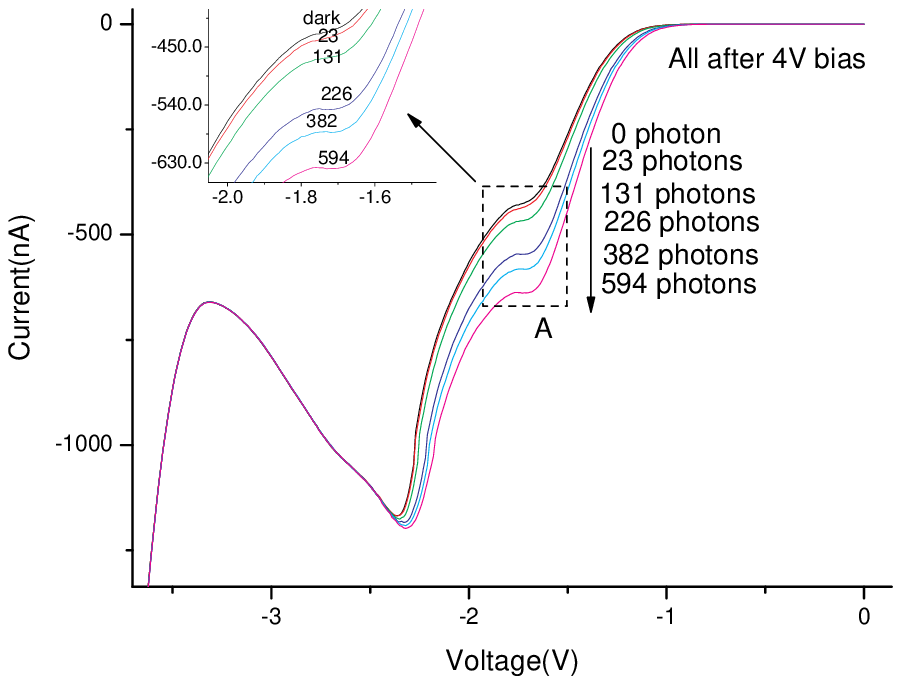}
  \caption{}
   \label{fig:Laser} 
\end{figure}

The electron charging condition of InAs dots can be modulated by photo-excited holes. Fig. \ref{fig:Laser} shows the \textit{I-V } characteristics of our device as a function of different light intensities of attenuated 532nm laser beam. A bias of 4V was applied to the device before the reverse sweep. Unlike the photon sensitivity observed in the previous AlGaAs QD-RTD,\cite{PRLphoton,APLphoton,APLphoton2} our structure shows higher sensitivity to the light irradiation under reverse bias than that in forward bias condition. Moreover, the voltage of peak A is insensitive to the light intensity. Fig. \ref{fig:Laser} shows the high responsivity of our device to low photon rates laser beam. With photon rate of 23 photons per second per $\mu m^{2}$, an increase of 10 nA in the current value of peak A is observed at 77K. This photon responsibility gives a multiplication factor in order of $10^{7}$ for photo-excited carrier in our QD-RTD device structure.

\begin{figure} 
  \centering 
  \includegraphics[width=\textwidth]{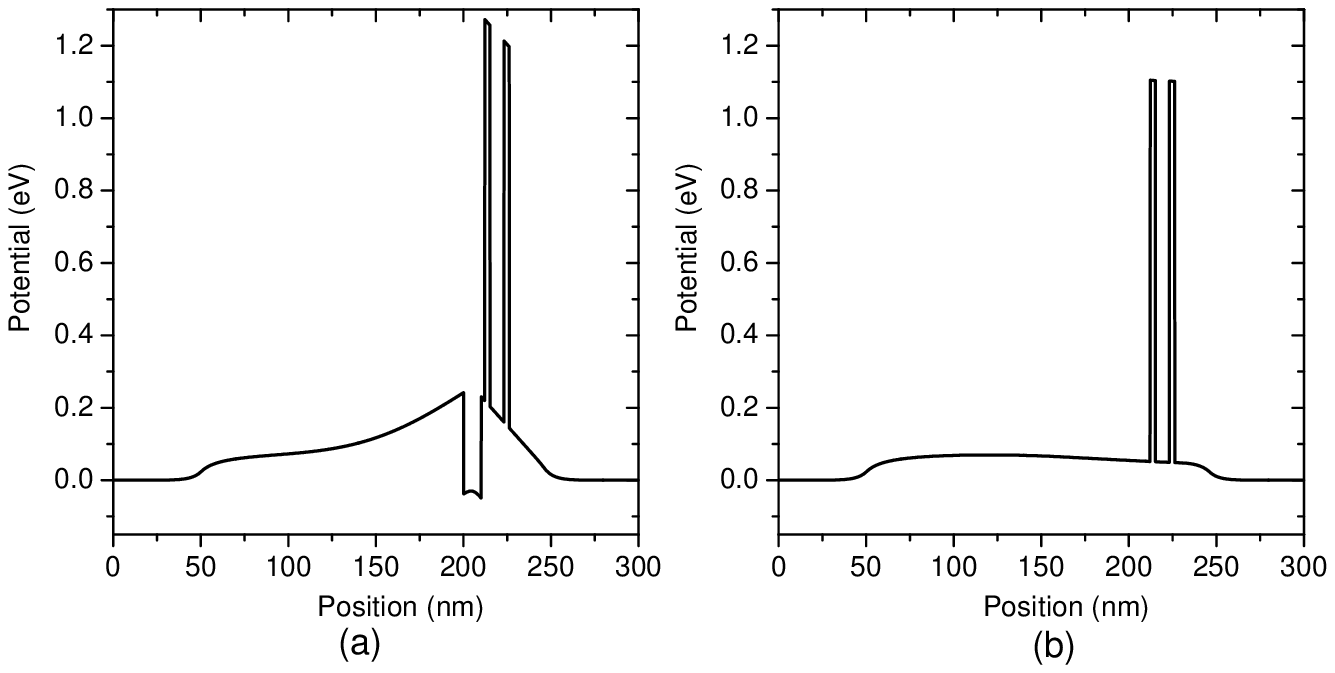}
  \caption{}
   \label{fig:QDRTDRTD} 
\end{figure}

\begin{figure} 
  \centering 
  \includegraphics[width=\textwidth]{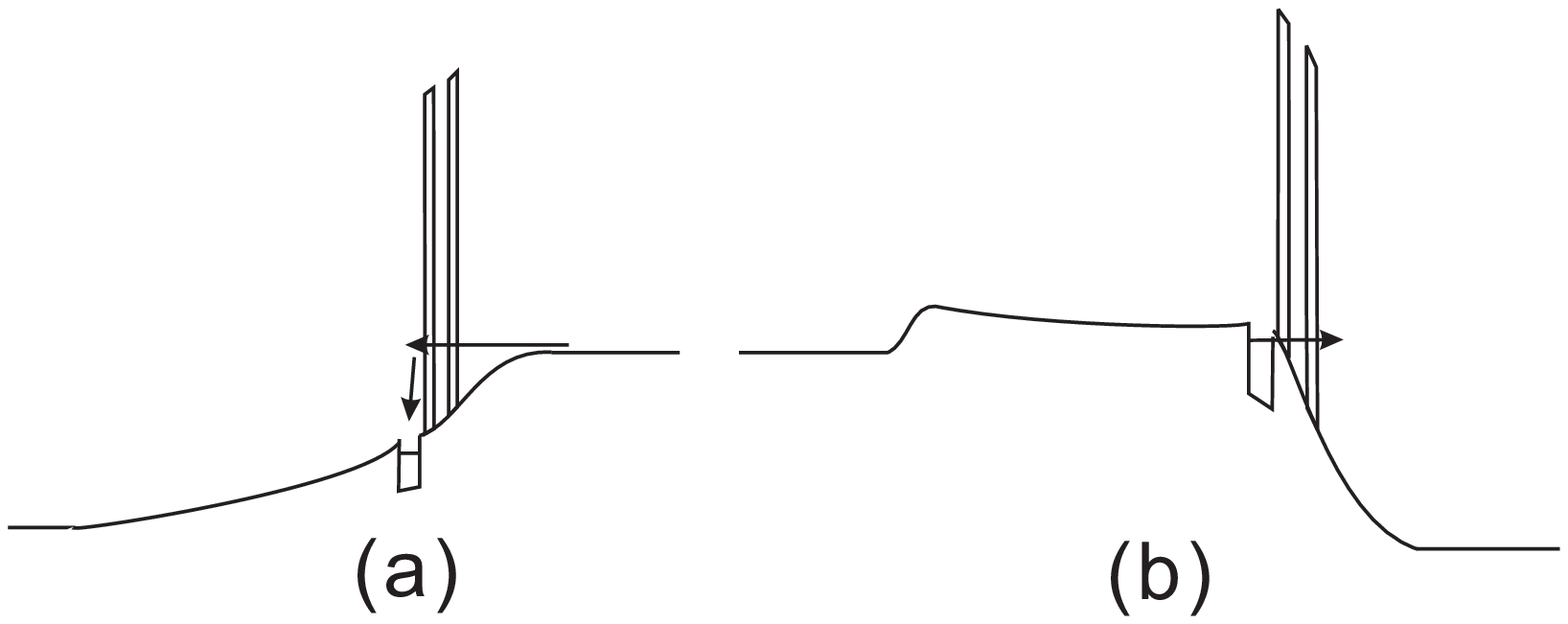}
  \caption{}
    \label{fig:explain} 
\end{figure}

The mechanism for the high multiplication factor for photo-excited carrier is presented in Fig. \ref{fig:QDRTDRTD}, Fig. \ref{fig:explain}. Fig. \ref{fig:QDRTDRTD} are band diagrams of QD-RTD structure in the case of with and without InAs dots, respectively, obtained by solving the Schr\"{o}dinger and Poisson equations self-consistently in the device.\cite{banddiagram} As an approximation, the InAs dot layer can be simply treated as the InAs quantum well for qualitative discussions. Fig. \ref{fig:QDRTDRTD} clearly indicates that the potential near charged dots is pushed up. When the InAs dots are heavily charged by the electrons due to the former forward bias, it will make the potential increase and thus make the main current flow through the region without InAs dots (see Fig. \ref{fig:AFMstructureIV}(a), Path$_{\textrm{RTD}}$). This resonant tunneling current has been observed as peak B and C in Fig. \ref{fig:AFMstructureIV}(b), resulted from electrons resonant tunneling from emitter through the quasi-bound states of the AlAs/GaAs/AlAs double barrier structure. The potential of the double barrier quantum structure area surrounded by the dots is also pushed up after forward bias, so that more reverse bias voltage should be needed to reach the resonance condition. This will make the absolute value of peak B voltage increase after former forward bias.

The peak A is resulted from the electrons escaping from InAs quantum dots via resonant states in the adjacent double barrier quantum structure. The charging and discharging of InAs QD are depicted in Fig. \ref{fig:explain}(a), (b) respectively. With the increase of the former forward bias, more electrons are charged into the InAs dots and the potential near the dots is pushed up accordingly before reverse bias sweep. This has the effect of reducing the electron current flowing through the dots along Path$_{\textrm{QD-RTD}}$ and thereby resulting in a weak peak A. The dots can be discharged through capturing the photo-excited holes and thus it will make the initial potential near the dots reduce gradually, which increases the electron current flowing along Path$_{\textrm{QD-RTD}}$. This photo-excited discharging process is also confirmed by the similarity between Fig. \ref{fig:AFMstructureIV}(b) and Fig. \ref{fig:Laser}. Since the resonant tunneling transition time is about 3 order shorter than the electron-hole recombination time in InAs QD,\cite{RTDtimeorder, dottimeorder} this allows a very high multiplication of photo-excited hole current by switching on the electron resonant tunneling current. So that one can get a high photon detection responsivity of QD-RTD in reverse bias.

In conclusion, we have demonstrated a way for high sensitive photon detection by QD-RTD in reverse bias with over charged InAs dots. The light sensitivity of our device in reverse bias was obtained as 10nA with incident photon rate of 23 photons per second per $\mu m^{2}$ on our 5 $\mu m^{2}$ device area at 77K. This may open more chance on technical application for the RTD based photon detection by uplifting the operation temperature from liquid helium (4.2K) to liquid nitrogen (77K).

\begin{acknowledgments}
This work was financially supported by State Key Basic Research Program of China (2006CB0L0607, 2006CB921204) and Grand Foundation of Shanghai Science and Technology (05DJ14003).
\end{acknowledgments}

\newpage

\begin{center}{\Large \bf  Figure Captions} \end{center}

\hfill

{\bf FIG.~1}: (a) Schematic of the device structure. The surface morphology of InAs dots is characterized by AFM. In reverse bias, top contact serves as emitter layer and two current pathes are presented as the lines labeled as Path$_{\textrm{RTD}}$ and Path$_{\textrm{QD-RTD}}$. Path$_{\textrm{QD-RTD}}$ passes through a InAs dot while Path$_{\textrm{RTD}}$ just passes the double barrier structure without dots. (b) Dark \textit{I-V} curves in reverse sweep (from 0V to -4V) at 77K as a function of different bias voltage before sweep. The former bias, in the range of -4V (bottom) to 4V (top), was applied to the device for about 2 seconds before the reverse sweep. Inset: Dark \textit{I-V} curve in forward sweep after -4V bias (botom) and 4V bias (top). The diagrammatic arrows in figure(a) indicate the origin of the observed peaks discussed in text. \vspace{1cm}

{\bf FIG.~2}: Reverse sweep \textit{I-V} curves at 77k as a function of different photon rate (per second per $\mu m^{2}$). Inset shows the enlargement of the rectangular part. \vspace{1cm}

{\bf FIG.~3}: Calculated energy band diagrams of the heterostructure at zero bias: (a) with InAs dots; (b) without InAs dots. \vspace{1cm}

{\bf FIG.~4}: Schematic band diagrams of QD-RTD: (a) InAs dots in charging mode (under forward bias); (b) InAs dots in discharging mode (under reverse bias). \vspace{1cm}

\newpage

\begin{thebibliography}{9}
\bibitem {forwardbias}The forward bias is defined as a positive voltage applied to the top contact of the device. The QD layer is in the collector side of the structure in forward bias.
\end{thebibliography}
\end{document}